\documentclass[aps,pre,reprint,superscriptaddress]{revtex4-1}
\usepackage{graphicx}
\usepackage[tight,centerlast]{subfigure}
\usepackage[english]{babel}

\draft 

\begin{document}


\title{Solid Surface Structure Affects Liquid Order at the Polystyrene/SAM Interface}


\author{Philipp Gutfreund}
\email{gutfreund@ill.eu}
\affiliation{Institut Laue-Langevin, 38042 Grenoble, France}
\affiliation{Condensed Matter Physics, Ruhr-University Bochum, 44780 Bochum, Germany}
\author{Oliver B\"aumchen}
\altaffiliation{Present address: Department of Physics \& Astronomy, McMaster University,  L8S 4M1 Hamilton, Canada}
\email{baumcho@mcmaster.ca}
\author{Renate Fetzer}
\altaffiliation{Present address: Karlsruhe Institute of Technology, Institute for Pulsed Power and Microwave Technology, 76344 Eggenstein-Leopoldshafen, Germany}
\affiliation{Department of Experimental Physics, Saarland University, 66041 Saarbr\"ucken, Germany}
\author{Dorothee van der Grinten}
\affiliation{Condensed Matter Physics, Ruhr-University Bochum, 44780 Bochum, Germany}
\author{Marco Maccarini}
\affiliation{Institut Laue-Langevin, 38042 Grenoble, France}
\author{Karin Jacobs}
\affiliation{Department of Experimental Physics, Saarland University, 66041 Saarbr\"ucken, Germany}
\author{Hartmut Zabel}
\author{Max Wolff}
\altaffiliation{Present address: Materials Science, Department of Physics and Astronomy, Uppsala University, 75121 Uppsala, Sweden}
\affiliation{Condensed Matter Physics, Ruhr-University Bochum, 44780 Bochum, Germany}


\date{\today}

\begin{abstract}
 We present a combined x-ray and neutron reflectivity study characterizing the interface between polystyrene (PS) and silanized surfaces. Motivated by the large difference in slip velocity of PS on top of dodecyl-trichlorosilane (DTS) and octadecyl-trichlorosilane (OTS) found in previous studies, these two systems were chosen for the present investigation. The results reveal
 the molecular conformation of PS on silanized silicon. Differences in the molecular tilt of OTS and DTS are replicated by the adjacent phenyl rings of the PS. We discuss our findings in terms of a potential link between the microscopic interfacial structure and dynamic properties of polymeric liquids at interfaces.
 \end{abstract}


\maketitle 

\section{Introduction}
When downsizing devices, confinement and interface effects grow enormously in importance. Apart from the fundamental interest in interfacial structure and dynamics, micro- and nanodevice fabrication open new perspectives for applications in manufacturing, pharmaceutics, chemistry or the food industry \cite{Squires2005,Thorsen2002,Rowland2008}. Especially, in the context of microfluidic devices \cite{Squires2005}, the controlled motion of small amounts of liquid is indispensable. 
As the solid/liquid friction dramatically impacts hydrodynamics in these systems, the boundary condition (BC) of flowing liquids, commonly quantified by the slip length \cite{Navier1823}, has been extensively revised on a microscopic length scale in recent years \cite{Neto2005,*Lauga2005,*Bocquet2007}.\\
Navier first supposed that a liquid may slip over a solid surface \cite{Navier1823} and introduced the slip length \textit{b}, which is defined as the distance \textit{z} from the interface where the velocity profile \textit{v(z)} of the liquid extrapolates to zero:
\begin{equation}
b=v(0)(\partial v(z)/\partial z)^{-1}|_{z=0}.
\end{equation}
The possibility of surface slippage was subsequently intensely discussed. Since the mid 19th century, the no-slip BC had been generally favored and in the 20th century, fluid dynamics textbooks assumed it in general, often without reference to its empirical origin. Experimental evidence of the failure of the no-slip BC was often attributed to parasitic effects or the lack of resolution, until de Gennes theoretically predicted large slip of entangled polymers in capillaries \cite{deGennes1979}. This provoked many investigations and today the slippage of entangled polymer melts is a well-known phenomenon \cite{Baeumchen2010}. 
In the late 1990s, slip of Newtonian liquids was observed in molecular dynamics (MD) simulations \cite{Barrat1999} and shortly after that it was experimentally confirmed \cite{Pit2000,Zhu2001}. Although interfacial slip developed to a well-recognized phenomenon, its microscopic origin is still unclear, also hindered by the fact that most experimental techniques used to determine the slip length are invasive or indirect.\\ 
From a theoretical point of view, two different types of slippage are distinguished \cite{Granick2003}: {\sl Real slip} occurs when the liquid slides over the solid surface on an atomic scale. 
Alternatively, {\sl apparent slip} arises where a microscopic boundary layer exists that is structurally and/or dynamically different from the bulk liquid. This boundary layer may lead to a different viscosity and is observed as interfacial slip on a larger length scale, although the no-slip BC may microscopically still hold. 
The nature of such a boundary layer may be a depleted density of the liquid \cite{Ruckenstein1983,*MaccariniSteitz2007} or an alignment of the near-surface molecules \cite{Heidenreich2007}.\\
Depletion effects of simple liquids 
have been observed in various cases \cite{Maccarini2007,Ocko2008,*Poynor2008,*Mezger2010,Chattopadhyay2010,*Mezger2011,*Chattopadhyay2011,Gutfreund2010} using x-ray and neutron reflectometry (XRR and NR). 
Their origin and the consequential link to macroscopic properties of liquids on solid surfaces such as hydrophobicity and also slippage is currently under debate \cite{Huang2008,Chattopadhyay2010,*Mezger2011,*Chattopadhyay2011,Gutfreund2010}. 
Density profile fluctuations and depletion layers of polymer melts close to solid substrates were also reported and attributed to altered molecular conformations and locally modified segmental distributions \cite{Bollinne1999}. In cases of entangled polymer melts, dedicated chain conformations at the solid/liquid interface are responsible for a decrease in the entanglement density compared to the bulk and, thus, substantially influence slippage \cite{Baeumchen2009}.\\
In this work, we present a combined XRR and NR study on polystyrene (PS) films on top of two different silanes, octadecyl-trichlorosilane (OTS) and dodecyl-trichlorosilane (DTS). As known from previous studies, PS melts show large slippage when flowing over hydrophobized surfaces, depending on molecular weight \cite{Baeumchen2009}, temperature and substrate \cite{Fetzer2005,*FetzerMuench2007,*Baeumchen2007}. 
In contact with DTS, a slip length of roughly 1\,$\mu$m (cf. Ref. \cite{Fetzer2005,*FetzerMuench2007,*Baeumchen2007}) has been found for PS of the same molecular weight (13.7\,kg/mol) and for the same annealing temperature (120\,$^{\circ}$C) used in this study. On OTS, however, the sip length is about one order of magnitude shorter for the same parameters. This strong effect on the slip length is surprising, as both are chemically identical self-assembled monolayers (SAM) that differ only by six backbone hydrocarbons in tail length. We show that the difference between the two surfaces is a lower grafting density of the DTS resulting in a tilt of the hydrocarbon tails. We provide evidence that the SAMs induce conformational changes within the interfacial polymer, which may influence slippage.\\

\section{Experimental Section}
The Si(100) wafers (Wacker/Siltronic, Burghausen, Germany, boron p-dotation, 10-20\,$\Omega$cm resistance) were hydrophobized with OTS and DTS monolayers \cite{Brzoska1994}, which resulted in 
a static contact angle of 67\,$\pm$\,3$^{\circ}$ for PS on the silanized wafers in both cases. The advancing water (Milli-Q synthesis system, Millipore, USA, organic impurities $<$\,6\,ppb, resistance at 25$^{\circ}$C: 18.2 M$\Omega$cm) contact angle was 116$^{\circ}$ on OTS and 114$^{\circ}$ on DTS. The receding contact angle was 110$^{\circ}$ in both cases. The atactic PS with a molecular weight of 13.7\,kg/mol ($M_{w}/M_{n}$=1.03) and the deuterated PS (\textit{d}PS) with a molecular weight of 12.3\,kg/mol ($M_{w}/M_{n}$=1.05) were purchased from PSS, Mainz, Germany. PS films between 50\,nm and 60\,nm were prepared by spin-casting a toluene (Merck, Darmstadt, Germany) solution onto mica and floating on Millipore water, from where they were picked up by the hydrophobized wafers. Then, the samples dedicated for the reflectivity experiments were annealed above the glass transition temperature ($T_{g}$) at 120\,$^{\circ}$C for 30\,s just before the onset of dewetting. Results of the dewetting studies and the slip length determination can be found elsewhere \cite{Fetzer2005,*FetzerMuench2007,*Baeumchen2007}. The bare silanized substrates were measured unannealed.\\
The x-ray measurements were conducted at beamline BL9 \cite{Krywka2006} of the Dortmund Electron Accelerator (DELTA), Germany, with photon energies of 11\,keV and 15.2\,keV and beam sizes of 0.2*2.5\,mm$^{2}$ and 0.1*1\,mm$^{2}$, respectively, with an angular resolution of 0.008$^{\circ}$ (FWHM). We observed no beam damage during the x-ray measurements. 
The NR measurements  were performed on the ADAM Reflectometer \cite{ADAM} at the Institut Laue-Langevin (ILL) in Grenoble, France, using a 0.5*10\,mm$^{2}$ beam with a constant angular resolution of 0.08$^{\circ}$ (FWHM). Pre-characterization of the samples dedicated for neutron scattering by means of XRR was done on the laboratory reflectometer XPert Pro PW3020, Panalytical, Netherlands. Fitting of the reflectivity data was obtained by using co-refinement of a slab model with \textit{Motofit} \cite{Motofit}. The error bars given by the fit correspond to one standard deviation, whereas the parameter ranges indicated in this work correspond to the variation among two measurements of equal samples from different batches.\\

\section{Results}
\subsection{X-ray reflectivity of bare silane substrates}
\begin{figure*}
\subfigure[]{\label{fig:DTSXrayRefl}\includegraphics[width=6.5 cm]{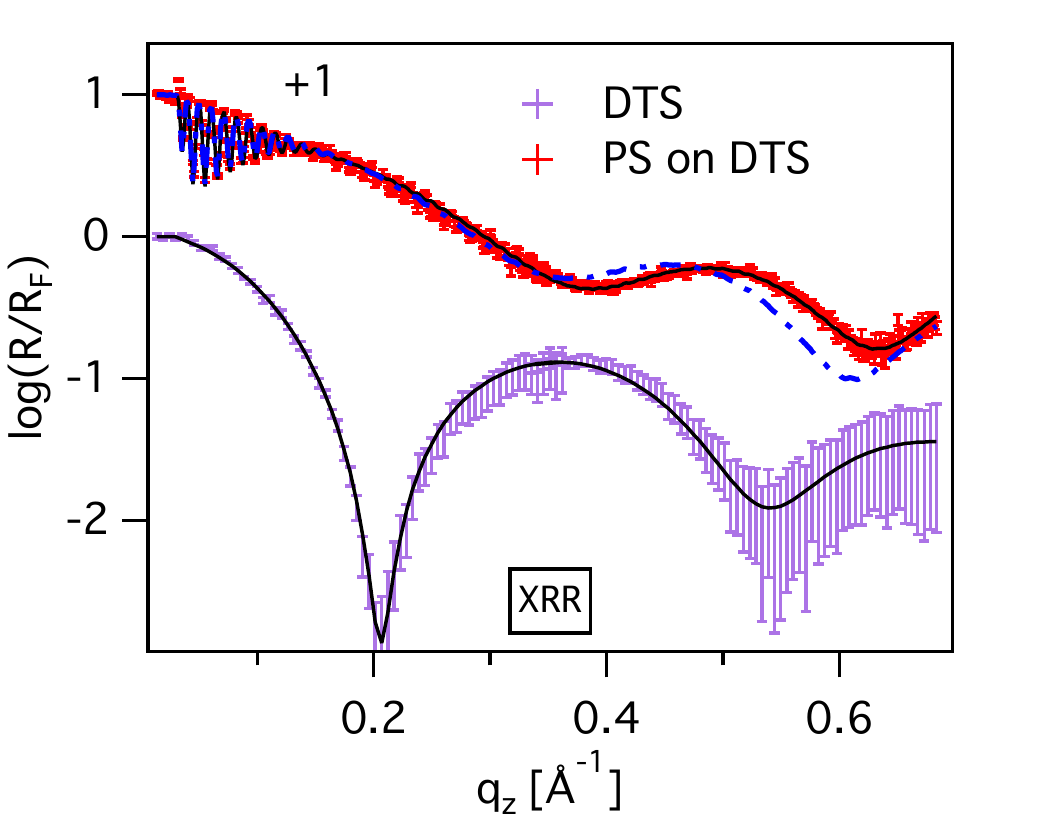}}
\subfigure[]{\label{fig:OTSXrayRefl}\includegraphics[width=6.5 cm]{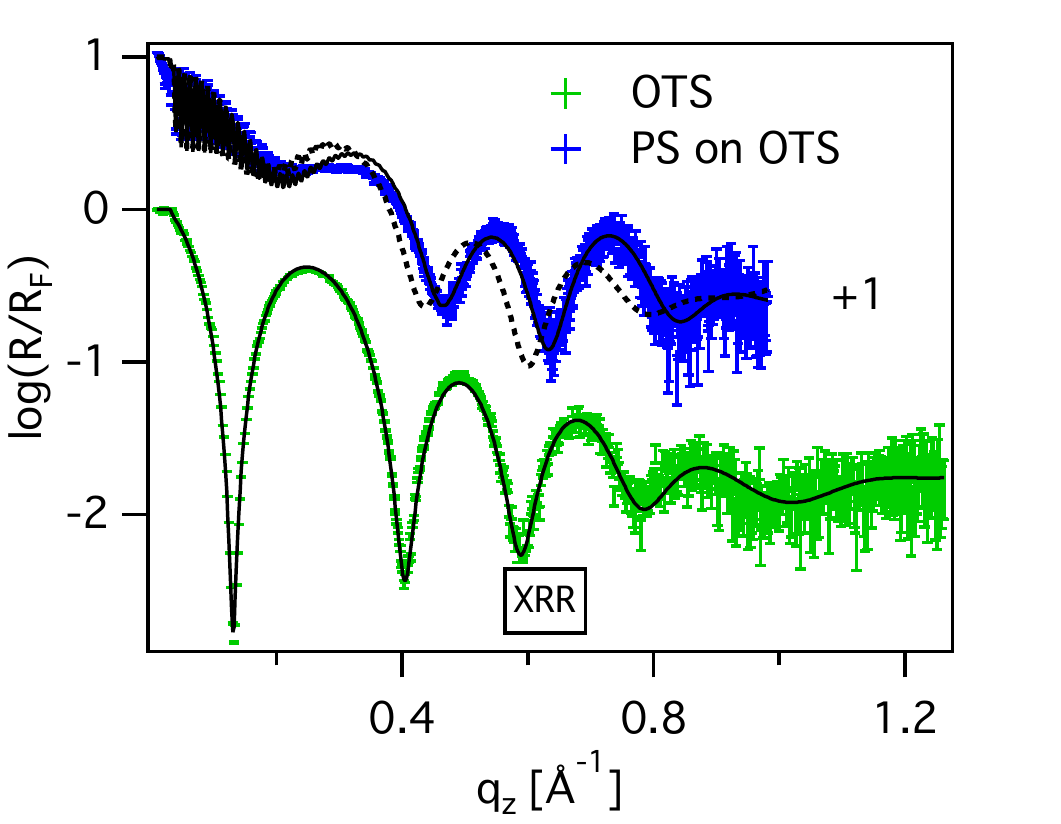}}\\
\subfigure[]{\label{fig:DTSXraySLD}\includegraphics[width=6.8 cm]{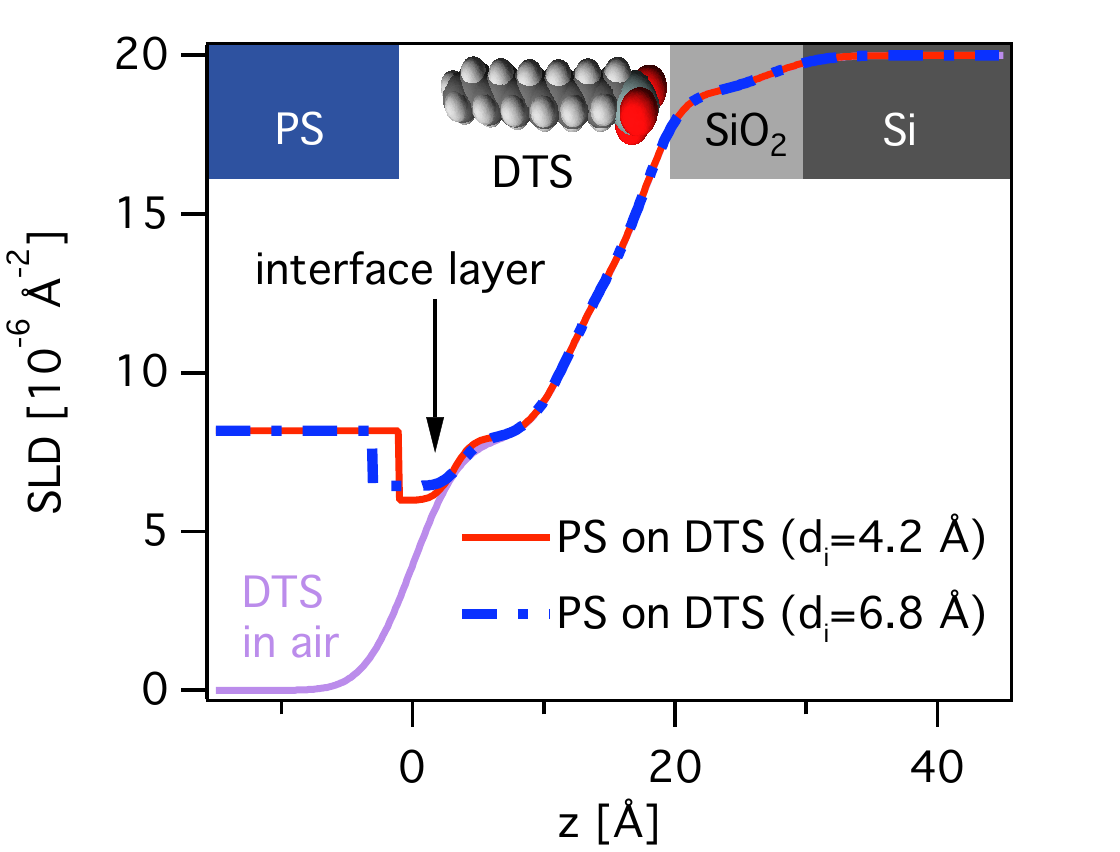}}
\subfigure[]{\label{fig:OTSXraySLD}\includegraphics[width=6.68 cm]{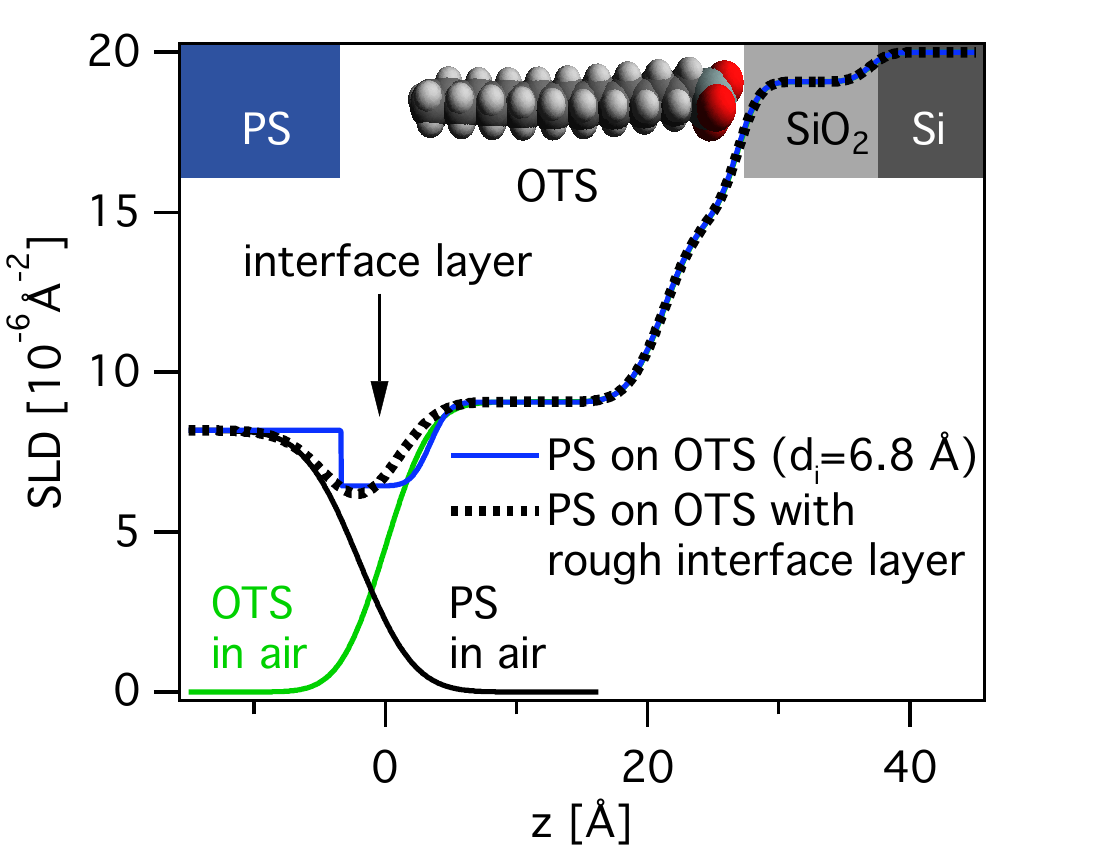}}
\caption{X-ray reflectivity curves normalized to the Fresnel-reflectivity on a logarithmic scale of the bare silicon wafers covered with DTS (purple bars) (a) and OTS (green bars) (b) as well as the reflectivities for the silanized samples covered with PS (shifted by one, red and blue bars, respectively). The solid lines represent fits. Fig.\ (c) and (d) display the corresponding scattering length density (SLD) profiles in the same color code as the reflectivity data. To exclude interfacial roughness as the origin of the depletion layer, the dotted black line (d) was obtained from adding the SLDs of the OTS and PS profiles in air. The resulting simulated reflectivity in (b) demonstrates that the experimental reflectivity data cannot be explained by just a roughness mismatch between OTS and PS. In order to highlight the difference in the depletion layer thickness for OTS and DTS, the blue dash-dotted lines in (a) and (c) denote a simulation of the DTS data with a thicker interfacial layer.} 
\label{fig:XrayRefl}
 \end{figure*}
 The x-ray reflectivities normalized to the Fresnel curve, \textit{i.e.}\ the reflectivity for an ideally flat silicon surface, are shown in Fig.\ \ref{fig:DTSXrayRefl} and (b). 
To get a quantitative description of the measured results, we first analyze the silanized Si wafers and assume a three-slab model, consisting of silicon oxide (SiO$_{2}$), a silane headgroup and a hydrocarbon tail \cite{Tidswell1990}. This reproduces the measured data, as can be seen by the solid lines in Fig.\ \ref{fig:XrayRefl}. The SAM parameters (see supplementary information) reflect the characteristics of completely grown silane layers \cite{Tidswell1990,Tidswell1991}: On top of the 9\,-\,10\,\AA\, thick SiO$_{2}$, with densities of 2.24\,-\,2.25\,g/cm$^{3}$, there is a silane headgroup, which is 5.6\,-\,5.95\,\AA\, thick, followed by the silane tail. The roughness between subsequent layers is between 1\,-\,4\,\AA . Note that the roughness of the bare DTS tail of $2.9\pm0.08$\,\AA\, is comparable to the roughness of $2.73\pm0.01$\,\AA\, measured for OTS. This rules out the possibility of surface roughness as a potential cause for difference in slippage between OTS and DTS. Moreover, in-plane rocking curves at different $q_{z}$ values on the reflectivity showed no significant broadening of the specular peak on both silanized surfaces which underlines the flatness and homogeneity of both samples. The only significant difference between the DTS and the OTS layer, apart from the tail length, is a higher grafting density of the OTS, which appears as a higher electron density of the silane head and tail. The headgroup of the DTS has an electron density of $0.476\pm0.007$\,\AA$^{-3}$, compared to $0.532\pm0.004$\,\AA$^{-3}$ in case of the OTS. Likewise, the density of the DTS tail ($0.82\pm0.01$\,g/cm$^{3}$) corresponds to 88\,\% of an alkane crystal's density \cite{Crissman1970}, whereas the OTS reaches 100\,\% ($0.936\pm0.004$\,g/cm$^{3}$). This difference is also observed when comparing the thicknesses of the layers with the calculated length of an all-trans hydrocarbon chain \cite{Tidswell1990}. The $21.31\pm0.05$\,\AA\, OTS tail length matches 99\,\% of the calculated fully stretched molecule (21.5\,\AA). The $12.0\pm0.1$\,\AA\, DTS tail length, however, corresponds to only 86\,\% of the calculated 13.9\,\AA, which is commonly explained as a randomly tilted SAM \cite{Tidswell1990,Bierbaum1995} and yields a tilt angle of 30$^{\circ}$ for DTS in this case. This difference between DTS and OTS is regularly observed \cite{Vallant1999} 
which may be due to the less optimal preparation temperature (room temperature) of DTS in comparison to the OTS \cite{Onclin2005}.\\ 
\subsection{X-ray reflectivity of PS on silane substrates}
When the silanized substrates are brought into contact with PS and annealed well above the bulk $T_{g}$ of PS, the data analysis reveals an interface layer with lower density between the SAM tail and the PS as can be seen in Fig.\ \ref{fig:XrayRefl}. To check whether this density-reduced layer is just an artifact of a roughness mismatch of the adjacent layers due to, for example, insufficient annealing we calculated the electron density profile of the interface between the OTS and PS layers by just adding both profiles measured in air before contact, assuming no change in the layers themselves. 
The simulated scattering length density (SLD) profile and the resulting reflectivity 
are denoted by black dotted lines in Fig.\ \ref{fig:XrayRefl}(b) and (d). This assumption does not reproduce the measured curve, obviously the PS has changed in contact with the SAM. 
The SLD profile corresponding to the best fit to the reflectivity data which is shown by the solid lines in the same figure, features a sharp density change of the interface layer towards the residual PS film which points out that a smooth interface has emerged on a molecular level. 
The relevant fit parameters are summarized in Tab.\ \ref{tab:Parameters} in the supplementary information. The major difference between the DTS and OTS interfaces is the thickness of the density-reduced interface layer. The $4.2\pm0.14$\,\AA\, thick layer at the PS/DTS interface is considerably thinner than the $6.79\pm0.04$\,\AA\, at the PS/OTS interface.\\ 

\subsection{Neutron reflectivity of PS on silane substrates}
\begin{figure}[]
 \includegraphics[width=7.5cm]{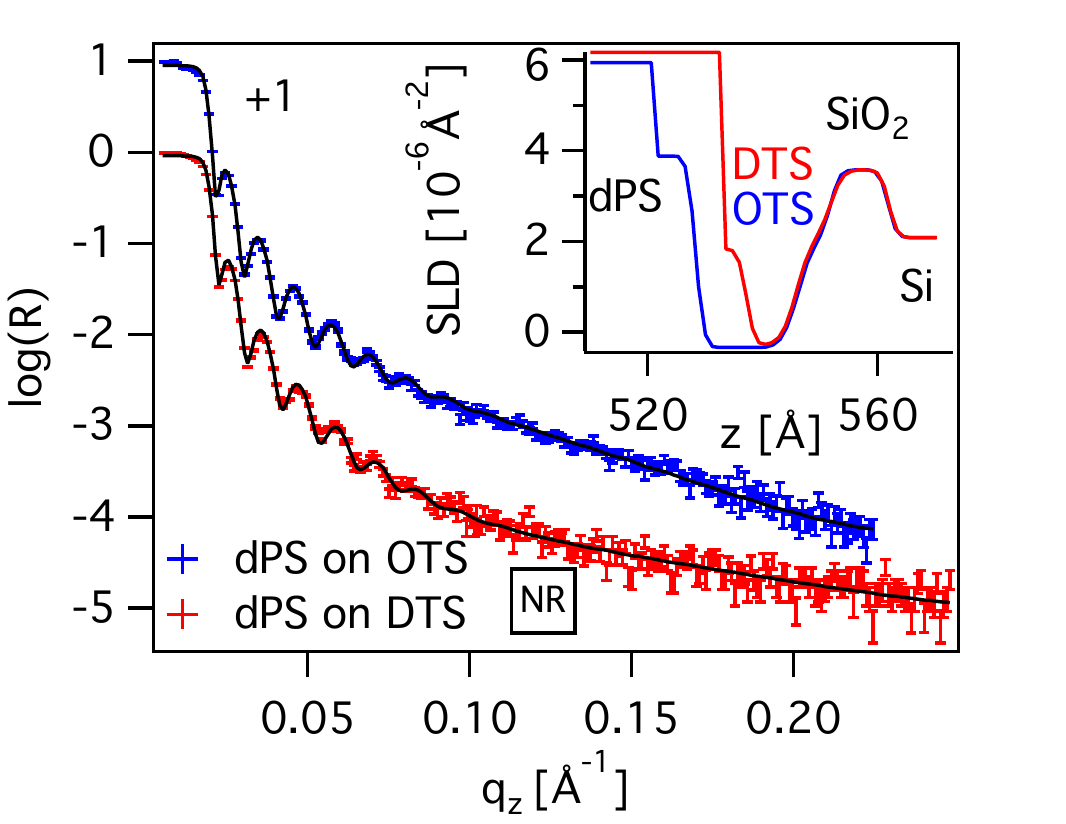}%
\caption{Neutron reflectivities in log scale of the OTS (blue bars, shifted by one) and DTS (red bars) substrates covered with deuterated PS. The solid lines are fits corresponding to the SLD profile from the inset in the same color code as the data points. \label{fig:NR}}
\end{figure}

In order to obtain information about the chemical composition of the interface layer between OTS/DTS and PS, we have performed NR experiments on the same system. In contrast to x-rays, which are sensitive to the electron density of the sample, neutrons are scattered by nuclei and the scattering length difference between a proton (-3.7\,fm) and a deuteron (6.7\,fm) is noticable. This makes NR particularly sensitive to protonated/deuterated interfaces.\\
Replacing the PS by $d$PS, we obtained the NRs shown in Fig.\ \ref{fig:NR}.
As the neutron measurements suffer from a considerably smaller momentum transfer range as compared to XRR their spatial resolution is much lower. Thus we fixed all fit parameters when analyzing the NR measurements by the corresponding x-ray values of the hydrogenated samples and varied only the SLD of the interfacial layer (and the SLD and thickness of the PS layer). For those parameters neutrons are more sensitive than x-rays due to the large contrast between protonated silane (SLD = -0.4\,*\,$10^{-6}$\AA$^{-2}$) and deuterated PS (SLD = 6.6\,*\,$10^{-6}$\AA$^{-2}$) (for x-rays the contrast is about 10 times smaller). This gives us a clear picture of the isotope composition of the interfacial layer. A purely protonated silane layer would result in a small negative SLD whereas a deuterated PS layer would show up in a layer with a SLD of about 4.5\,*\,$10^{-6}$\AA$^{-2}$.\\ 
In the inset of Fig.\ \ref{fig:NR} it is evident that none of the two scenarios is observed for the SLD of the interfacial layer. Instead both interfaces seem to consist of both protonated and deuterated material. By taking the electron density from the x-ray SLD and the nuclear density of the NR SLD the exact amount of PS and Silane of the density-reduced layer can be calculated. For the OTS interface, this results in $66\pm4\,\%$ PS and $11\pm3\,\%$ silane as compared to their bulk density, and, in case of the DTS, $32\pm12\,\%$ PS and $43\pm12\,\%$ silane are present. This means that the observed low-density layer in the x-ray measurements comprises parts of the SAM and the adjacent PS. 
This result is in-line with recent x-ray reflectometry studies of water at hydrophobic surfaces \cite{Ocko2008,*Poynor2008,*Mezger2010,Chattopadhyay2010,*Mezger2011,*Chattopadhyay2011} and clarifies that the density-depleted liquid often observed at solid/liquid interfaces is partly due to the presence of the hydrogen termination of the hydrophobic SAM whose SLD is close to 0 for x-rays.\\ 

\section{Discussion}
\begin{figure}
\subfigure[]{\label{fig:OTSPSSketch}\includegraphics[width=6 cm]{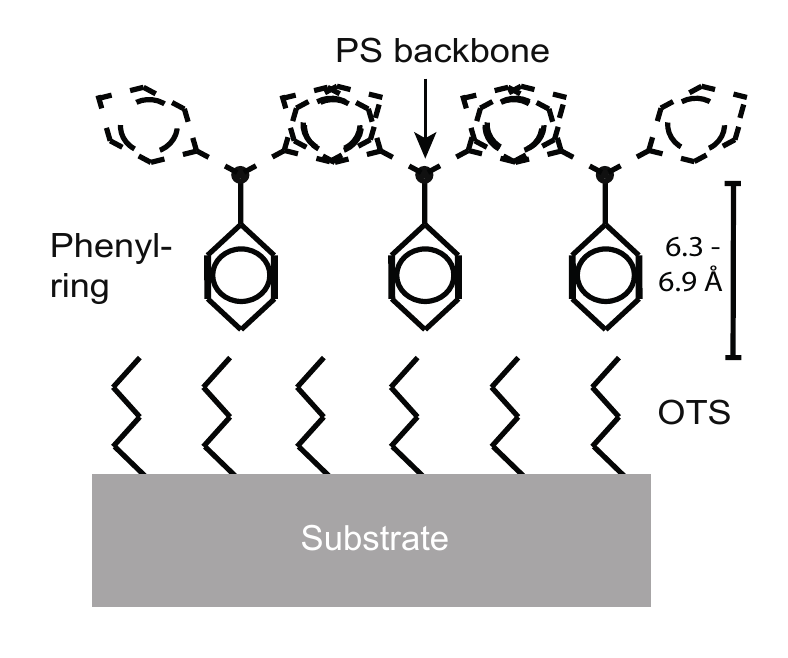}}\\
\hspace{1.3 cm}\subfigure[]{\label{fig:DTSPSSketch}\includegraphics[width=6.5 cm]{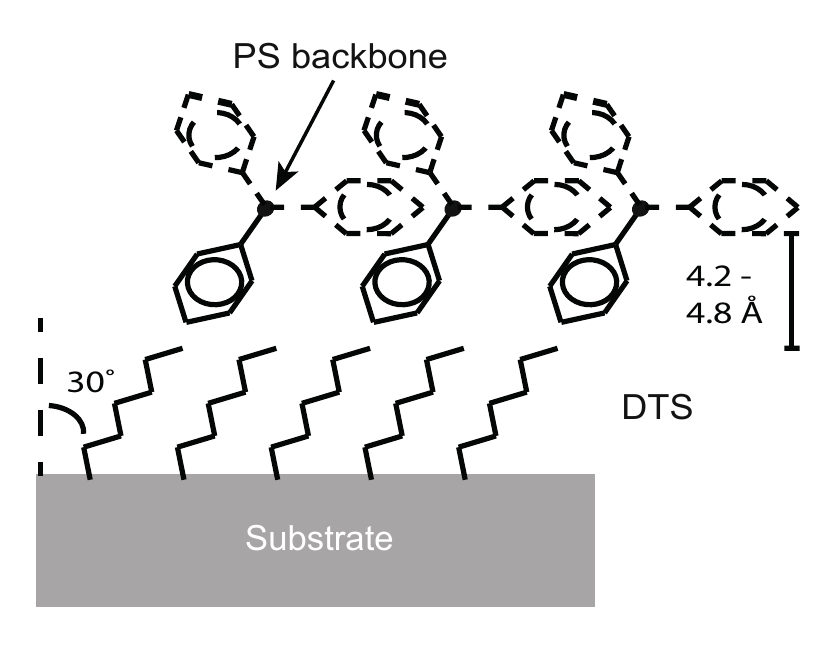}}
\caption{Sketch of the molecular conformation at the PS/OTS (a) and PS/DTS (b) interface as explained in the text.}
\label{fig:SilanePSSketch}
 \end{figure}

To produce the sharp density step of PS in contact with the SAMs, as seen by the electron density profile
, the adjacent PS chains cannot be randomly oriented. Instead, we assume that a rather well-ordered arrangement of contacting chain segments is formed. Considering the molecular composition of PS, only an orientation with the phenyl groups pointing to the SAM complies with all parameters extracted from the scattering experiments. This scenario is sketched in Fig~\ref{fig:OTSPSSketch}.
The distance between the hydrocarbon backbone of the PS and the end of the phenyl group, including the covalent radius of the hydrogen, is 5.6\,\AA. The projected bonding length of the OTS hydrogen termination including its covalent radius is 0.65\,-\,1.25\,\AA, depending on whether the covalent radius of the linked carbon is subtracted or not. In total this adds up to 6.25\,-\,6.85\,\AA, in accordance with the 6.8\,\AA\, thick interface layer observed at the PS/OTS interface. Due to three possible orientations of the phenyl group around a flat PS backbone, in average only every third one would be incorporated in the interface layer. In the bulk, one and a half out of three phenyl rings are projected on one side of the PS backbone. This explains the density of deuterated material, reduced by one third, in the interface layer revealed by NR. If we additionally assume that the phenyl rings follow the orientation of the SAM, they would be tilted by 30$^{\circ}$ as depicted in Fig.\ \ref{fig:DTSPSSketch} in contact with DTS. This would lead to a reduced interface layer of 4.18\,-\,4.78\,\AA, which matches very well with the 4.2\,\AA\, thick interface layer observed at DTS. 
In contrast to one methyl-hydrogen pointing into the interface layer at OTS, two methyl-hydrogens are present in the interface layer of the tilted DTS. Additionally, the interface layer at DTS is thinner than the OTS one and hence, the proportional amount of protonated silane should be considerably higher at the DTS, at the expense of the relative amount of deuterated PS. This is well in accord with the 43\,\% silane and 32\,\% $d$PS as revealed by NR for the DTS interface.\\ 
The interfacial order which we deduce from our scattering experiments is in line with recent MD simulations, where a crystalline surface was able to induce order in polymeric liquids \cite{Priezjev2010}: The first liquid layer showed an almost perfect reproduction of the preset periodic crystal structure. Recent experiments confirm, that in thin polymer films chain segments may order \cite{Rivillon2000} and, in particular, that certain orientations of the PS phenyl rings can be induced by the presence of interfaces \cite{Mukhopadhyay2010}, even in absence of specific interactions: As demonstrated by non-linear optical techniques such as sum-frequency generation (SFG) spectroscopy, the interplay of intra- and intermolecular interactions causes the phenyl rings to point away from the bulk polymer film perpendicular to the polymer/air interface \cite{Gautam2000} and also towards a hydrophobic substrate \cite{Wilson2002}.\\
Concerning slippage, MD studies report slip lengths on the order of several monomer lengths \cite{Priezjev2010} which are quantitatively not comparable to the large experimental slip lengths (up to several $\mu$m close to $T_{g}$) for unentangled polymer melts on silanized surfaces \cite{Fetzer2005,*FetzerMuench2007,*Baeumchen2007}. Experimentally determined slip lengths may incorporate apparent slip in addition to real slip. Apparent slip of polymeric liquids may be explained by a higher segmental mobility in the vicinity of the interface, either (i) due to an interfacial depletion effect \cite{Ruckenstein1983} or (ii) due to layering \cite{Barrat1999,Thompson1990,*Priezjev2004,*Zhu2004} and/or alignment \cite{Heidenreich2007} of the liquid near the interface. As a third potential mechanism (iii), a reduced segmental friction coefficient between the adjacent polymer chains and the substrate may be caused by particular polymer conformations near interfaces. Recent MD simulations on slipping oligomers \cite{Vadakkepatt2011} highlight that the most significant part of the energy transfer (friction) between the solid and the liquid is dissipated in the first liquid layer and only minor energy transfers occur between subsequent layers.\\
The structural data of our study imply a flat arrangement of the adjacent PS chain segments and a sharp step in the density profile between the interfacial layer and the residual polymer film. The density was shown to be reduced down to 75 - 77\,\% in a depletion zone of 4 - 7\,\AA\, thickness. Mechanism (i) does not apply to our system since the small extent of the depletion zone cannot account for the large experimental slip lengths measured. Regarding mechanism (ii), apparent slip due to layering implies more than one layer of polymer to be aligned. However, density oscillations indicating molecular layering were not detected. Such shear-induced effects have been observed in \textit{in situ} experiments for a different system while applying shear flow \cite{WolffMagerl2004}. Our experiments clearly demonstrate a distinct orientation of the phenyl rings of the PS melt due to the structural properties of the adjacent substrate (OTS and DTS). Hence, compared to the non-oriented bulk liquid, the presence of locally deviating dynamical properties such as friction and viscosity in the interfacial region are very likely.\\
Theoretical studies report a suppression of slip for a higher degree of substrate-induced liquid ordering near a flat smooth surface (c.f. \cite{Priezjev2006} and referenes therein). This is qualitatively in line with the difference in slip length that has been experimentally observed for PS on OTS and DTS. In general, an ultra-low surface roughness of the substrate, as it is found in the case of OTS and DTS, is a precondition for the presence of ordering effects. In the limit of commensurability of spatial dimensions of the surface corrugations and the molecular size of the liquid, a strong suppression of slip can be achieved \cite{Priezjev2006}. This might explain small slip length values (between 0 and 100\,nm at maximum) that have been experimentally found for the same PS on an amorphous substrate exhibiting slightly larger surface roughness \cite{Baeumchen2009}. Studies of the interface structure of the latter system are on-going.\\
In literature, experimentally observed differences in slip length of Newtonian liquids on different substrates have been widely attributed to surface properties such as roughness and the strength of interaction between liquid molecules and the substrate \cite{Pit2000,Zhu2002,*Leger2003,*Schmatko2006}. We stress the fact that these parameters are found to be identical for PS on DTS and on OTS. For these systems, we provide evidence of a molecular interplay of the interfacial structure of the liquid and the surface order of the solid, that might affect a macroscopically detectable parameter, namely the amount of slip of a PS film on silanized surfaces.\\

\section{Summary and Outlook}
In summary, we have revealed that surface structure of a self-assembled monolayer affects the conformation of polymer chain segments adjacent to the solid boundary. The results of combined x-ray and neutron scattering studies point out that the adjacent polystyrene chain segments lie completely flat and, moreover, that the orientation of the phenyl rings replicates the self-assembled monolayer structure. Both facts appear to be the clue to understand a) substantially different polymer slippage on silanized surfaces exhibiting identical surface energies and polystyrene contact angles and b) large effective (comprising real and apparent) slip, both observed experimentally. Our findings corroborate on-going research claiming conformational changes at the interface in case of entangled polymeric liquids \cite{Baeumchen2009} and the interfacial liquid structure in case of non-entangled oligomers as important parameters governing macroscopic slip \cite{Schmatko2005,Gutfreund2010}. 
Additionally, our results might also shed light on further interfacial phenomena such as depletion layers or glass-transition temperatures of thin polymer films. Since the orientation of the phenyl rings of the polystyrene is linked to the aforementioned ordering phenomena, molecular dynamics simulations investigating slippage of polymer melts should intend to account for their entire monomeric structure to achieve full comparability to experimental situations.\\


\begin{acknowledgements}
 We thank Michael Paulus and Christian Sternemann for the x-ray beam time on DELTA and we gratefully acknowledge financial support by the BMBF (05K10PC1), the DFG grants ZA161/18 and JA905/3 within the priority program (SPP) 1164 and the graduate school GRK1276.
\end{acknowledgements}


%

\newpage

\begin{widetext}
\section*{Supplementary}
\label{sec:Supplementary}
\begin{table}[h!]
\caption{Fitting parameters of the silane tail and the interfacial layer at the SAM/PS interface measured with X-ray reflectometry.  \textit{d}, $N_{b}$ and $\sigma$ correspond to the thickness, scattering length density and roughness of the respective layers.\label{tab:Parameters}}
\begin{center}
  \begin{tabular}{lllll}
   \hline
   \hline
   Parameter & DTS in air & OTS in air & PS on DTS & PS on OTS\\
   \hline
   $N_{b}^{Si}$\,[$10^{-6}$\AA$^{-2}$] & $20\pm0.1$ & $20\pm0.1$ & $20\pm0.1$ & $20\pm0.1$\\
   $\sigma^{Si/SiO_{2}}$\,[\AA] & $3.3\pm0.8$ & $1.1\pm0.1$& $3.3\pm0.8$& $1.1\pm0.1$\\
   $d^{SiO_{2}}$\,[\AA] & $9\pm1$ & $10\pm0.1$& $9\pm1$& $10\pm0.1$\\
   $N_{b}^{SiO_{2}}$\,[$10^{-6}$\AA$^{-2}$] & $18.8\pm0.3$ & $19.08\pm0.05$& $18.8\pm0.3$& $19.08\pm0.05$\\
   $\sigma^{SiO_{2}/head}$\,[\AA] & $2.1\pm0.5$ & $1.4\pm0.1$ & $2.1\pm0.5$ & $1.4\pm0.1$\\
   $d^{head}$\,[\AA] & $5.93\pm0.02$ & $5.62\pm0.02$ & $5.93\pm0.02$ & $5.62\pm0.02$\\
   $N_{b}^{head}$\,[$10^{-6}$\AA$^{-2}$] & $13.4\pm0.2$ & $15\pm0.1$ & $13.4\pm0.2$ & $15\pm0.1$\\
   $\sigma^{head/tail}$\,[\AA] & $2.5\pm0.2$ & $2.19\pm0.04$ & $2.5\pm0.2$ & $2.19\pm0.04$\\
   $d^{tail}$\,[\AA] & $12\pm0.1$ & $21.31\pm0.05$ & $8.84\pm0.08$ & $17.86\pm0.03$\\
   $N_{b}^{tail}$\,[$10^{-6}$\AA$^{-2}$] & $7.9\pm0.1$ & $9.07\pm0.02$ & $7.9\pm0.1$ & $9.07\pm0.02$\\
   $\sigma^{tail/air}$\,[\AA] & $2.9\pm0.08$ & $2.73\pm0.01$&&\\
   $\sigma^{tail/interf}$\,[\AA] &&& 1.14 (fixed) & $1.14\pm0.03$\\
   $d^{interf}$\,[\AA] &&& $4.19\pm0.14$ & $6.79\pm0.04$\\
   $N_{b}^{interf}$\,[$10^{-6}$\AA$^{-2}$] &&& $5.99\pm0.03$ & $6.44\pm0.03$\\
   $\sigma^{interf/PS}$\,[\AA] &&& 0 (fixed) & $0\pm0.2$\\
   $d^{PS}$\,[\AA] &&& $539.9\pm0.6$ & $615.3\pm0.5$\\
   $N_{b}^{PS}$\,[$10^{-6}$\AA$^{-2}$] & & & $8.2\pm0.5$ & $8.2\pm0.5$\\
   $\sigma^{PS/air}$\,[\AA] &&& $3.2\pm0.02$ & $3.31\pm0.04$\\
     \hline
     \hline
     \end{tabular}
     \end{center}
\end{table}
  \end{widetext}

\end{document}